\documentclass[final,5p,times,13p]{elsarticle}
\usepackage{graphicx}
\usepackage{color}
\usepackage{amssymb}
\usepackage{amsmath}
\usepackage[T1]{fontenc}
\usepackage[colorlinks,bookmarks=false,citecolor=blue,linkcolor=red,urlcolor=black]{hyperref}
\begin{document}

\title{ Optimized Configuration Interaction Approach for Trapped Multiparticle Systems Interacting Via Contact Forces  
}


\author{Przemys{\l}aw Ko\'{s}cik}
\ead{koscik@ujk.edu.pl}

\address{Institute of Physics, Jan Kochanowski University, ul. \'{S}wi\c{e}tokrzyska 15, 25-406 Kielce, Poland}

\begin{abstract}

For one-dimensional systems with delta-contact interactions,  the convergence of the exact-diagonalization method is tested with a basis of  harmonic oscillator eigenfunctions with  frequency  $\Omega$  optimized
through the minimization
of the eigenenergy of the desired level. It is shown that  within the framework of this approach the  well-converged results can be  achieved  at much smaller  dimensions of the  Hamiltonian matrix   than with  the standard approach that uses $\Omega=1$. We present calculations for model systems of identical bosons with harmonic and double-well potentials. Our results show promising potential for diminishing the computational cost of numerical simulations of various systems of trapped ultracold atoms.\\

\end{abstract}


\maketitle

\section{Introduction}Over the past two decades, there has been an explosion of interest in the properties of
one-dimensional (1D) quantum gases with short-range interactions modelled by a Dirac $\delta$-potential.
 With the emergence of new technologies and experimental techniques, the properties of these systems, such as the number of particles, their interactions-, and the shape of the trapping potentials, can be controlled with high accuracy
\cite{kauf,serwane}. As a result, it has become possible to simulate various physical phenomena in controllable ways that can even provide the opportunity to realize experimentally different toy models.
Various physical realizations of systems of cold atoms
are nowadays achieved \cite{kauf,serwane,Greiner,Selim2012,catani,Selim2013,Selim2015}.
In view of this tremendous technological progress, research activity has exploded in the area of investigating the many-body properties of various quantum composite cold-atom systems
  \cite{Sowinski2016a,Zinner2017,Zinner2016,Har,Diaz2017,frank,bosebose,Koscik2018,sow,sow1,sow2}.

There are few 1D systems with contact interactions that can be solved analytically. The best known of these is the system composed of two identical particles held together in a harmonic trap, which has closed-form solutions in the whole interacting regime   \cite{Busch1999}.
The Bethe ansatz method is known to be a solution for the 1D Bose gas in the absence of an external potential; that is, the so-called Lieb-Liniger model  \cite{LL}. The most important result
  regarding the 1D gas   is the   famous      Bose-Fermi mapping  theorem \cite{Girardeau} that maps the Tonks-Girardeau (TG) gas of bosons with infinitely strong repulsions to a  free-fermion gas, which  does not depend on the external potential.
     As a result, the theoretical study  of   TG gases is a relatively easy task even in the limit of large particle numbers.
  The first observation of  TG gases in experiments \cite{Kinoshita}  provided theoretical communities with the impetus to study the properties of TG gases under different external potentials \cite{Xi,Murphy1,Murphy,periodic}. The Bose-Fermi mapping theorem also provides a tool for studying properties of multicomponent mixtures of strongly interacting gases \cite{mix0,mix1,mix2,mix3}.
  It is worth  pointing out  that a powerful pair-correlated variational approach to studying the ground states of  bosonic systems with  a  harmonic trap has been developed in \cite{Brouzos} and subsequently extended to  fermionic mixtures \cite{Brouzos1} and  to bosonic systems with different interactions
between the pairs of atoms \cite{barf}. This approach can also be extended   to other   mixtures of  cold atoms, such as boson-fermion mixtures \cite{mix0} and  bosonic mixtures \cite{Pyzh}.

However, in most cases, full numerical calculations are required to describe the transition  between the weakly and strongly correlated regimes, and this is generally a cumbersome task. Numerical simulations of  ultra-cold gases are often performed with the exact-diagonalization (ED) method and in the framework of the multiconfiguration time-dependent Hartree  method  \cite{MCTDH} that has been extended  for  bosons \cite{Mbosons} and fermions  \cite{Mfermions}, as well as  for bosonic (fermionic) mixtures \cite{MLMCTDHX1,MLMCTDHX2}.

The standard ED method is based on the Rayleigh-Ritz procedure and uses as a variational wave function a finite linear combination of  many-particle states of a proper symmetry under the exchange of particles, usually made up of solutions of the corresponding one-particle system. In contrast to variational methods that use single-trial wave functions, which are usually specialized to treat only ground states of specific systems, the ED method enables precise determination of ground- and higher bound-states in a systematic way.

In particular, most of the studies  available in the literature about 1D systems with harmonic trapping potentials $x^2/2$ use the ED method with harmonic oscillator (HO) eigenfunctions:
\begin{equation}\label{basis}
u_{n}(x)=\left( {\sqrt{\Omega}\over \sqrt{\pi} 2^n n!} \right)^{{1\over 2}}\mathrm{e}^{-{\Omega x^2\over 2}}\mathbf{H}_n\left(\sqrt{\Omega}
   x\right),
\end{equation}
with $\Omega=1$, where $\mathbf{H}_n$ is the $n$th order Hermite polynomial. However,
this results in  very poor convergence of the many-body eigenstates  as a function of the number of basis states \cite{pol}.
In fact, even for systems with  small particle numbers,  huge numbers of many-particle functions are needed to describe strongly interacting regime \cite{frank}.

In this paper we show that the ED method with basis functions given by (\ref{basis}) can be  a very effective tool for studying various trapped  systems with delta  interactions,  provided  the parameter $\Omega$ is variationally optimized.
The structure of this paper is as follows. Section \ref{tb}
outlines the formalism of the optimized ED (OED) approach. Section \ref{results} tests the convergence  of the OED method  for the examples of harmonic and  double-well potentials. Specifically,  a significant improvement in the  convergence  is demonstrated for
the harmonic trapping potential compared to the case without optimization of $\Omega$ ($\Omega=1$).
 Finally, section \ref{con} presents some concluding remarks.

\section{Optimized ED Approach
}\label{tb}  Without loss of generality,  we  deal only with   systems of identical  bosons.
We begin with the dimensionless Hamiltonian to deal with any confining potential: \begin{equation}\label{Hamiltonian_total}
 \hat{{\cal H}} = \sum_{i=1}^N h_{0}(x_{i}) + g\sum_{i<j}\delta(x_i-x_j),
\end{equation}
where the strength of
the interaction is governed by the coefficient $g$ and  $h_{0}$ is the one-body Hamiltonian given by \begin{equation}\label{one particle Hamiltonian}
h_{0}(x)=-\frac{1}{2}\frac{\partial^2}{\partial x^2} + { V}(x).
\end{equation}

  The true $N$-particle bosonic wave function can be
  represented as a linear combination:
\begin{equation}\label{exp}
|\Psi\rangle=\sum_{\beta}a_{\beta} |\textbf{u}_{\beta}\rangle
.\end{equation}
Here, $|\textbf{u}_{\beta}\rangle$ denotes the permanents that are constructed from the
one-particle basis (\ref{basis}), which  in the   occupation-number representation take  the form \begin{equation}\label{manyparticle}
|\mathbf{u}_{\beta}\rangle=|n_{0},n_{1} ,...\rangle_{\Omega}.
\end{equation}
This represents the fact that  the one-particle state $|i\rangle$ is occupied $n_{i}$ times, $\sum_{i}n_{i}=N$, and   $\beta$ labels the different distributions of the particles.
A feature  worth stressing  here is that   if the trapping potential $V$ is symmetric in $x$, then the corresponding
Hamiltonian (\ref{Hamiltonian_total}) commutes with the  symmetry operator $\hat{{\cal P}}$ defined as
$\hat{{\cal P}}\Psi(\mathbf{r})= \Psi(-\mathbf{r})$,
 the eigenvalues of which are  $p=\pm1$, $\mathbf{r}=(x_{1},x_{2},...,x_{N})$. Consequently,
the states with  parities $p=1$ and $p=-1$ are  superpositions of even-
and odd-parity  permanents, $\sum_{i} i n_{i}$= (even  or  odd) respectively.

In practical calculations, we must truncate  the many-particle basis.
One  reliable  way of doing this is  to use  the  basis made up of  Fock  states in the form
    $|\mathbf{u}_{\beta}^{K}\rangle=|n_{0},n_{1} ,..., n_{K},0,0...\rangle_{\Omega}$, where $\sum_{i=0}^{K} i n_{i}<K$ \cite{frank,shel,shel1,pl}. From now on  $D$ denotes the number of many-body basis functions that compose the truncated basis set.
Diagonalization    of the corresponding truncated  Hamiltonian matrix $[{H}_{\alpha\beta}]$, with ${H}_{\alpha\beta}=\langle{\mathbf{u}}_{\alpha}^K| \hat{{\cal H}}|{\mathbf{u}}_{\beta}^K\rangle$, thus yields a set of approximations to the energies,  $E_{i}^{(K)}$,  and the corresponding  eigenvectors $a^{(K)}_{i}$:
\begin{equation}\label{exp1}
|\Psi\rangle_{i}\approx\sum_{\beta}(a_{i}^{(K)})_{\beta} |\mathbf{u}_{\beta}^{K}\rangle
.\end{equation}
 By
increasing $K$,  approximations
to a larger and larger number of states are obtained with systematically
improved accuracy. However, the truncation of the basis set makes the    approximate eigenstates  dependent   on $\Omega$. Only in the limit as $K$ tends to infinity does  the dependency on $\Omega$ vanish altogether. This freedom in choosing the value of $\Omega$  can be used to improve the convergence \cite{okop,saad}.
Following  the principle of minimal sensitivity \cite{pms}, the parameter   $\Omega$ should be chosen
 so that the  approximations to a given physical quantity are as minimally  sensitive to its  variations  as  possible. Clearly,  the best   approximation of the $K$th order to  the   energy of  the desired state  is obtained for the value of $\Omega$ at which the corresponding eigenenergy $E_{i}^{(K)}$ attains its   minimum, i.e., $dE_{i}^{(K)}/
d\Omega|_{\Omega=\Omega_{opt}}=0$.
For large truncation orders,   finding  $\Omega_{opt}$ requires diagonalization of the truncated   Hamiltonian matrix   many times for different values of $\Omega$ until the minimum of $E_{i}^{(K)}$ is found.
It is worth mentioning that various  strategies for  fixing  the value of $\Omega$  before  diagonalization of the  Hamiltonian matrix  have been  tested on  single-particle systems (see \cite{Koscik2009} and reference therein for an overview).  However, none of these strategies   guarantees that
 the desired state will  be  estimated with optimal  accuracy.

   Here, we concentrate on testing the effectiveness of the strategy based on the minimization of $ E_{i}^{(K)}$.
    Although, strictly speaking, this approach yields the best approximation of the  $K$th order only to
     the considered energy level, the corresponding resulting wave function is usually determined with an
     accuracy that is close to the optimal one.
    \section{Results}\label{results}  For testing the performance of the OED scheme, we first choose the ground states of particles subjected to a harmonic confining potential. Since  the ground state is an even-parity state ($p=1$),
  the dimensions of  the   Hamiltonian matrix  can be reduced by including  in the calculations
   only the   permanents that satisfy  $\sum_{i=0}^{K} i n_{i}$=(even), which considerably diminishes  the
computational cost. In our calculations, the numerical minimization of $ E_{0}^{(K)}$ is done in  the framework of   Newton`s  iterative method for finding roots.
To illustrate this clearly,    we present in Table  \ref{tab:table1} an example of results of the first few iterations obtained in the $N=3$ case.

\begin{figure}
\begin{center}
\includegraphics[width=0.236\textwidth]{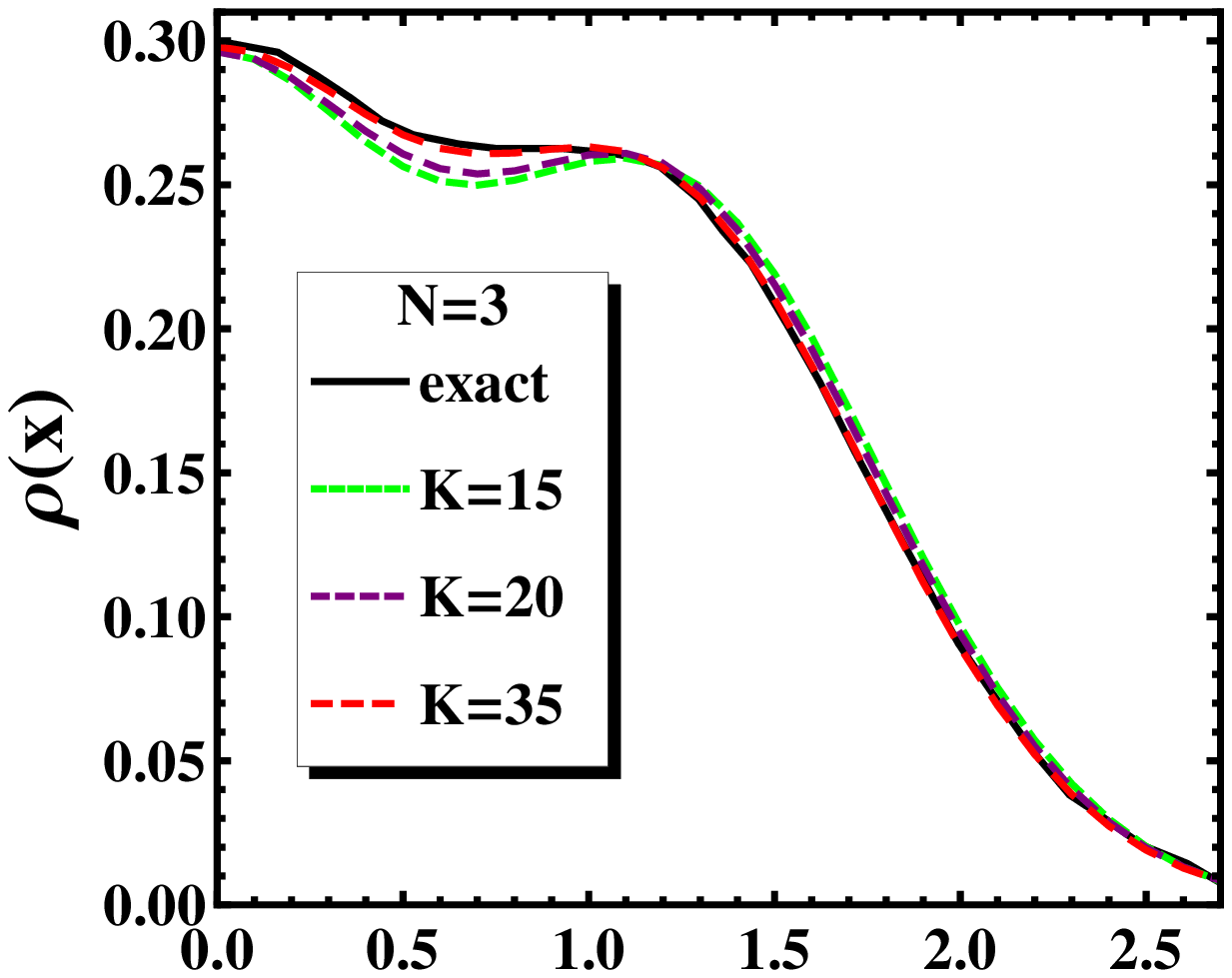}
\includegraphics[width=0.240\textwidth]{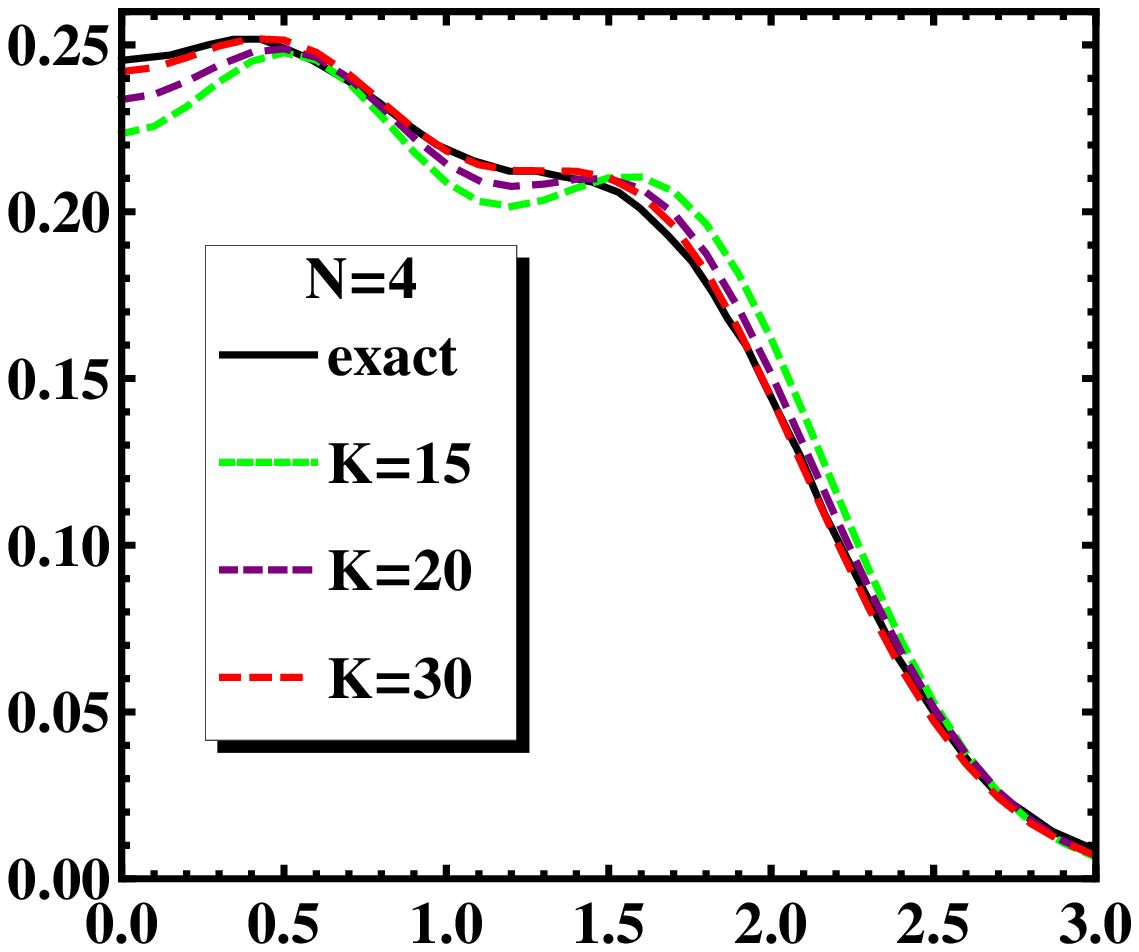}

\includegraphics[width=0.236\textwidth]{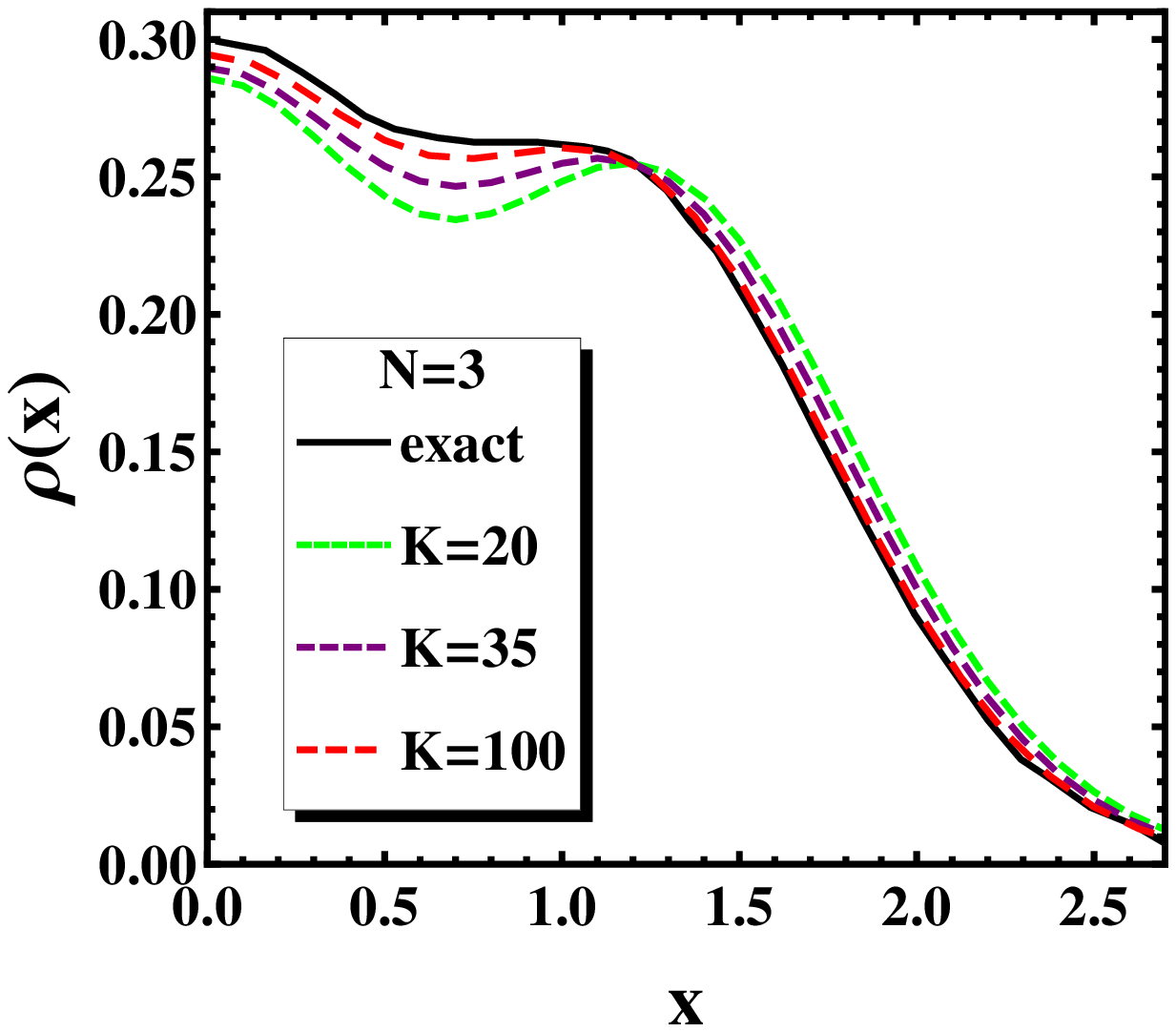}
\includegraphics[width=0.240\textwidth]{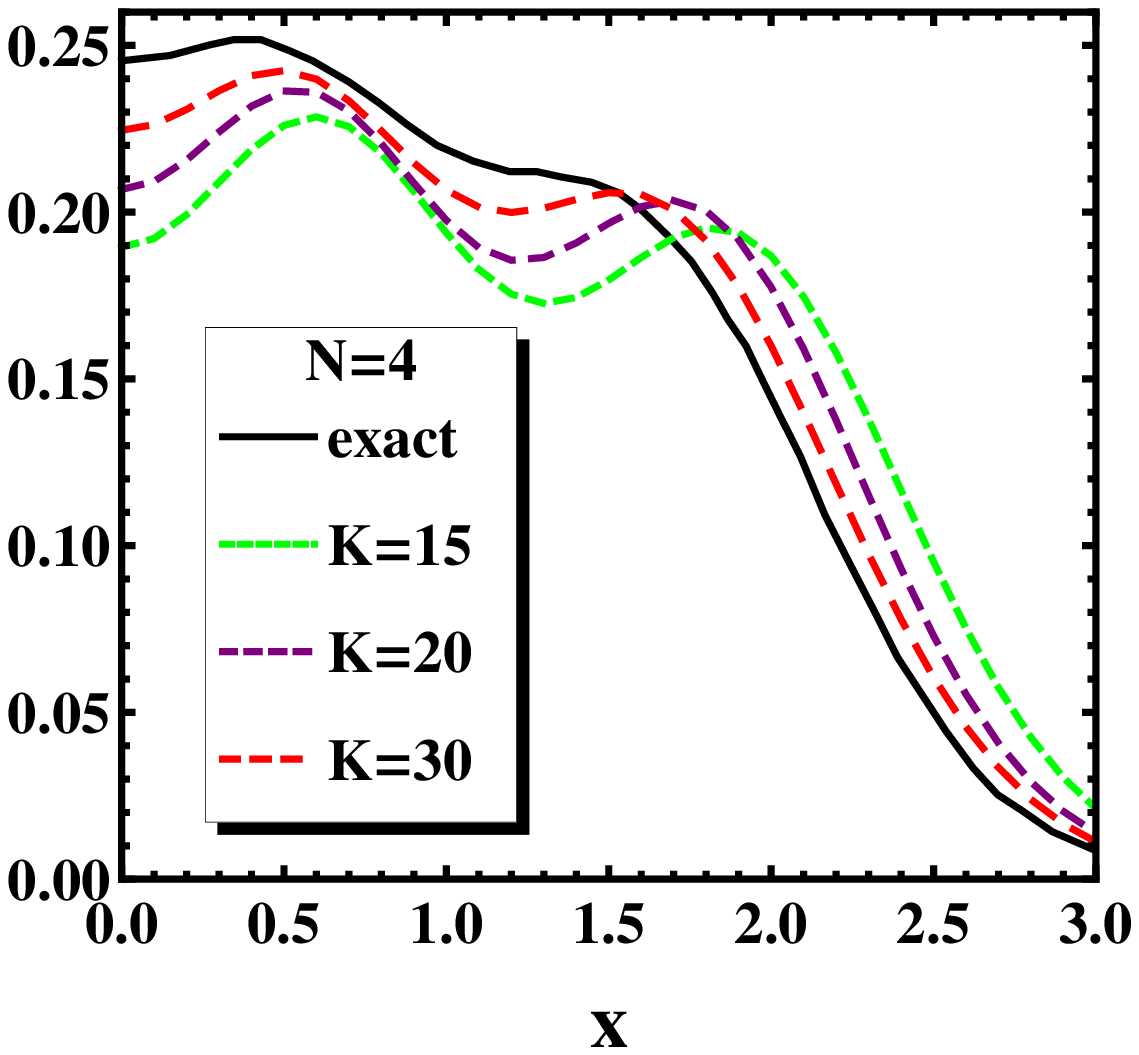}
\end{center}
\caption{\label{Fig1}
Profiles of one-body densities obtained for $N=3,4$ at $g=10$ for different cut-off values of $K$. Black continuous lines mark the  reference  results taken from  \cite{Brouzos} (Figure 2 (b-c) in this reference). The top-panel results were obtained through  the optimization strategy based on the minimization of  $E_{0}^{(K)}$. The corresponding numbers   $D$ and the optimal values of $\Omega_{opt}$  are as follows:
$\mathbf{N=3}\mathbf{:}K = 15~ (D= 81,\Omega_{opt} =2.59)\mathbf{;}
K = 20 ~ (D= 148, \Omega_{opt} =3.13)\mathbf{;}
K = 35 ~(D=762 ,\Omega_{opt} =5.11)\mathbf{;}\mathbf{N=4}\mathbf{:}K = 15~ (D=136, \Omega_{opt} =1.88)\mathbf{;}
K = 20 ~(D= 284, \Omega_{opt} =2.33)\mathbf{;}
K = 30 ~ (D= 1152, \Omega_{opt} =3.38$). The bottom-panel results  were obtained at $\Omega=1$.
}
\end{figure}

\begin{figure}[h]
\begin{center}
\includegraphics[width=0.239\textwidth]{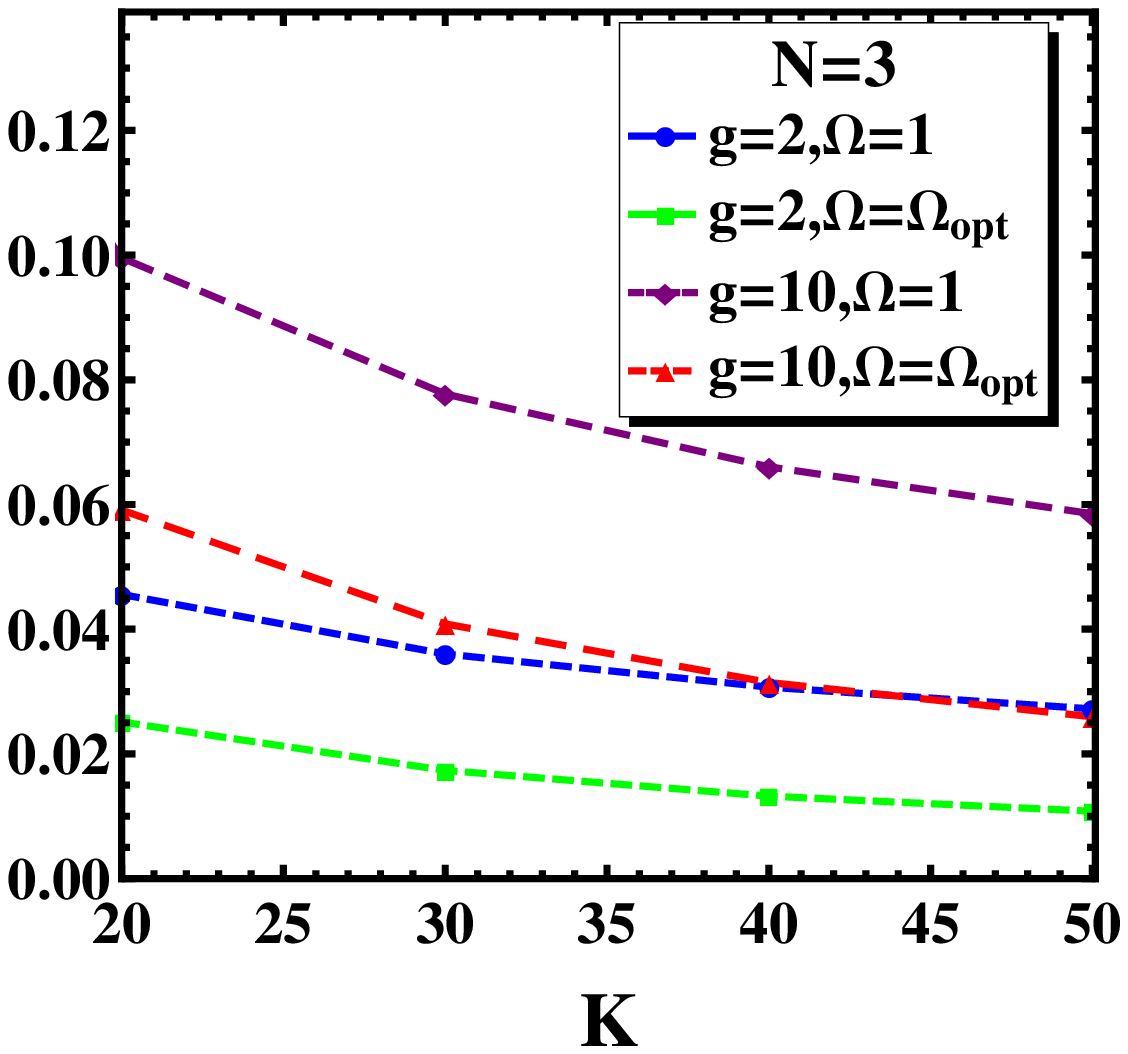}
\includegraphics[width=0.239\textwidth]{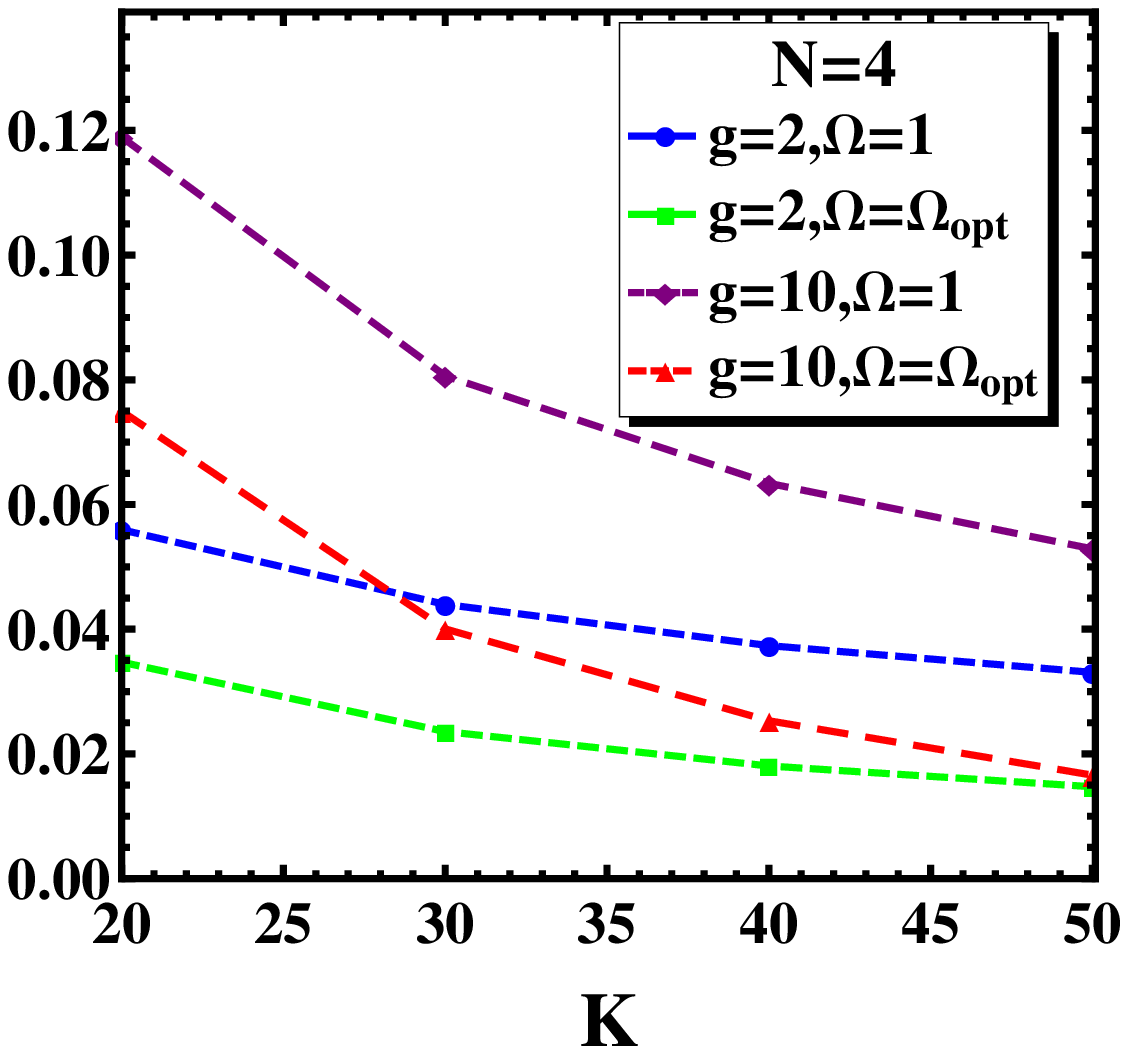}
\end{center}
\caption{\label{Fig2}
Relative deviations $\delta E=(E_{0}^{(K)}(\Omega)-E_{exact})/E_{exact}$  for ground states of  three-and four- particle systems, calculated at $\Omega=1$  and at  $\Omega=\Omega_{opt}$ for two different interaction strengths  $g=2$ and $g=10$, as  functions of the cut-off $K$. We have taken  the reference energies $E_{exact}$ from  Figure 2a in  \cite{Brouzos}.}
\end{figure}

 Here, we take as the reference  points  the  results
   obtained  for three- and four- particle systems in  \cite{Brouzos}, where  these have been determined in different ways with satisfactory accuracies.
In Figure \ref{Fig1} we present  the   one-body densities \begin{equation}\rho(x)=\int_{\Re^{N-1}}|\Psi(x,x_2,...,x_{N})|^2dx_{2}...dx_{N},
\end{equation}  obtained from the ED  calculations  before
 and after optimization of $\Omega$, for  different  cut-off values of $K$ at the strong interaction strength
$g = 10$, in the bottom and top panels, respectively. Black  continuous lines mark the  reference densities taken from  \cite{Brouzos}.  For the sake of completeness,  we   give  the optimal
values of $\Omega$ in the caption to this figure.
   We also calculated  the deviations of the  approximate energies  $E_{0}^{(K)}$  from the \textit{exact} energies, as functions of $K$,
 both  with and without optimization of $\Omega$.   Our results for  three- and four-particle calculations  at two transparent values of $g$, including $g = 10$, are displayed in Figure \ref{Fig2}.

\begin{table}[h]
\begin{center}
\begin{tabular}{lllll}
\hline
$\Omega_{0}=1$&$...$&$\Omega_{5}\approx5.15$&$\Omega_{6}\approx5.11$&$\Omega_{7}\approx5.11$ \\
\hline
$ 4.26943$&$...$ & $4.12803 $&$4.12802 $ &$4.12802$\\
\hline
\end{tabular}
\caption{\label{tab:table1}  Applying Newton`s method to the present problem yields the following recurrence equation for the term  $\Omega_{n+1}$:   $\Omega_{n+1}=\Omega_{n}- e^{(1)}_{n}/e^{(2)}_{n}$, where   $ e^{(1)}_{n}$ and $ e^{(2)}_{n}$ are  finite difference approximations to  first and second derivatives of $E_{0}^{(K)}$ at $\Omega=\Omega_{n}$, which are calculated here with a step length of  $d\Omega=0.005$. The table presents the   results of the first few Newton iterations  $\{\Omega_{n}, E_{0}^{(K)}(\Omega_{n})\} $  obtained    for the ground-state of the three-particle  system with  $g=10$ at  $K=35$.  In spite of the fact that the starting point differs considerably from an optimal solution, a fast convergence is observed.}
\end{center}
\vspace{-0.6cm}
\end{table}

 It is apparent from our results that there are clear benefits to optimizing the ED method.
We can conclude from the caption to the figure \ref{Fig1}   that the rate of convergence
 strongly depends on $\Omega$. In fact, the optimal value  $\Omega_{opt}$ increases  with the increase in  $K$.
 As might be expected,   it is clear that approximations to the energies are better after applying the optimization. This effect is more pronounced  when the
parameters of $g$ enter the strong-correlation regime; that is,   where the true wave functions have  more complex  structures.
However, the improvement of the  convergence is more pronounced in  one-body densities.  For example,  the  \textit{exact}  three-particle density is well reproduced in the OED calculation when $K$ is about three times smaller than  in the case where optimization is absent; that is, $K=35~(D=765)$ vs. $K=100~(D=14739)$.
It should be emphasized that the calculation time in the former case (iterative minimization) was considerably  shorter than in the latter one.

 The optimization of $ \Omega $ is therefore  crucial for reducing  the number of basis functions needed for accurate numerical calculations, as well as for  shortening the computation time. In general, the size of the computations that can be done depends not only on the efficiency of the numerical calculation software that is used but also on the computer system itself, especially on the size of  its memory. Our analysis clearly shows that the  OED method consumes less memory  than the standard ED approach ($\Omega=1$), and this makes it possible for systems with larger numbers of particles to be treated. Here, we leave an open question regarding how many particles can be treated effectively by modern supercomputers when using the OED approach. However, this number can be expected to be much larger than in the cases where there is no optimization of $\Omega$.

\begin{figure}
\begin{center}
\includegraphics[width=0.239\textwidth]{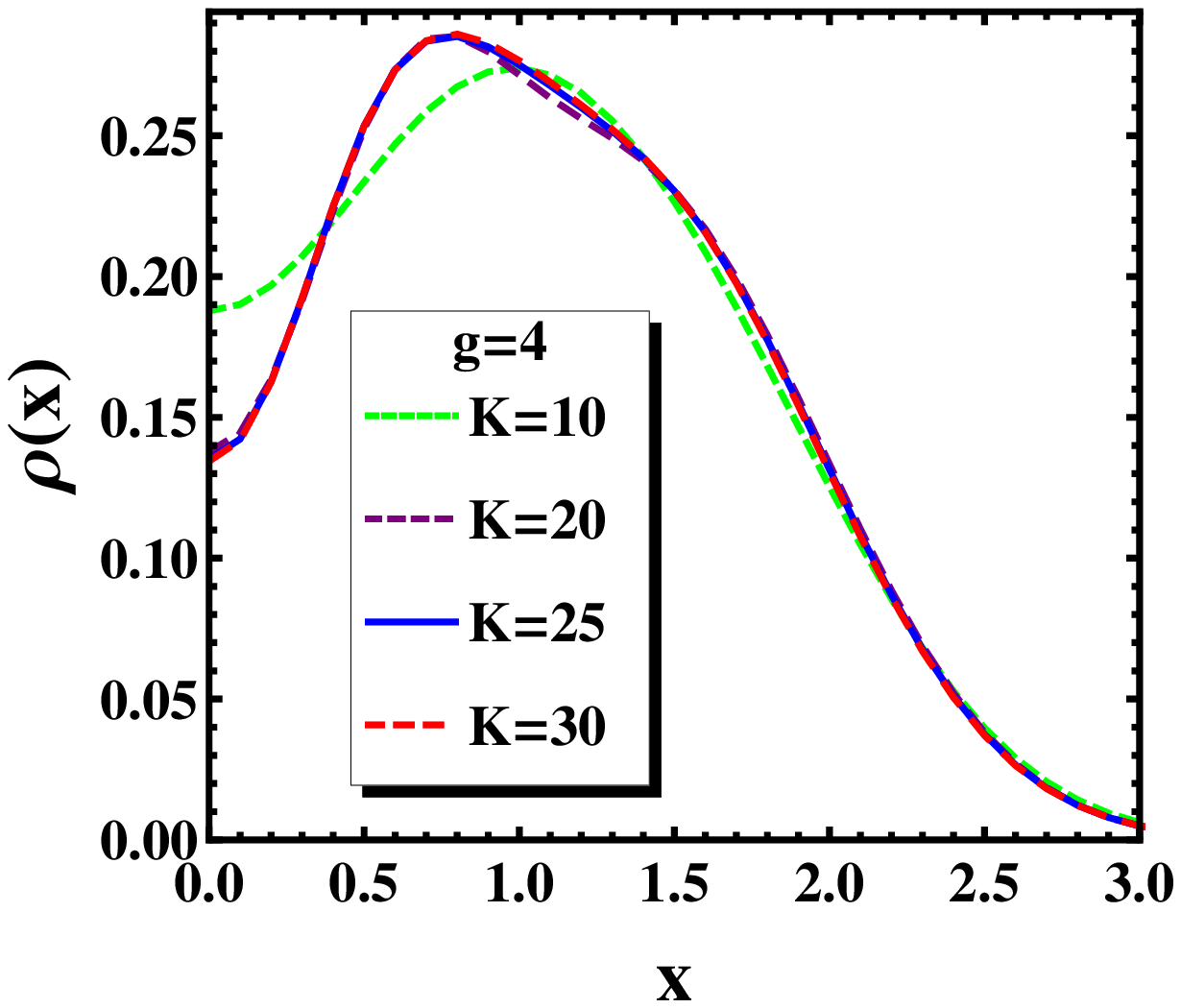}
\includegraphics[width=0.233\textwidth]{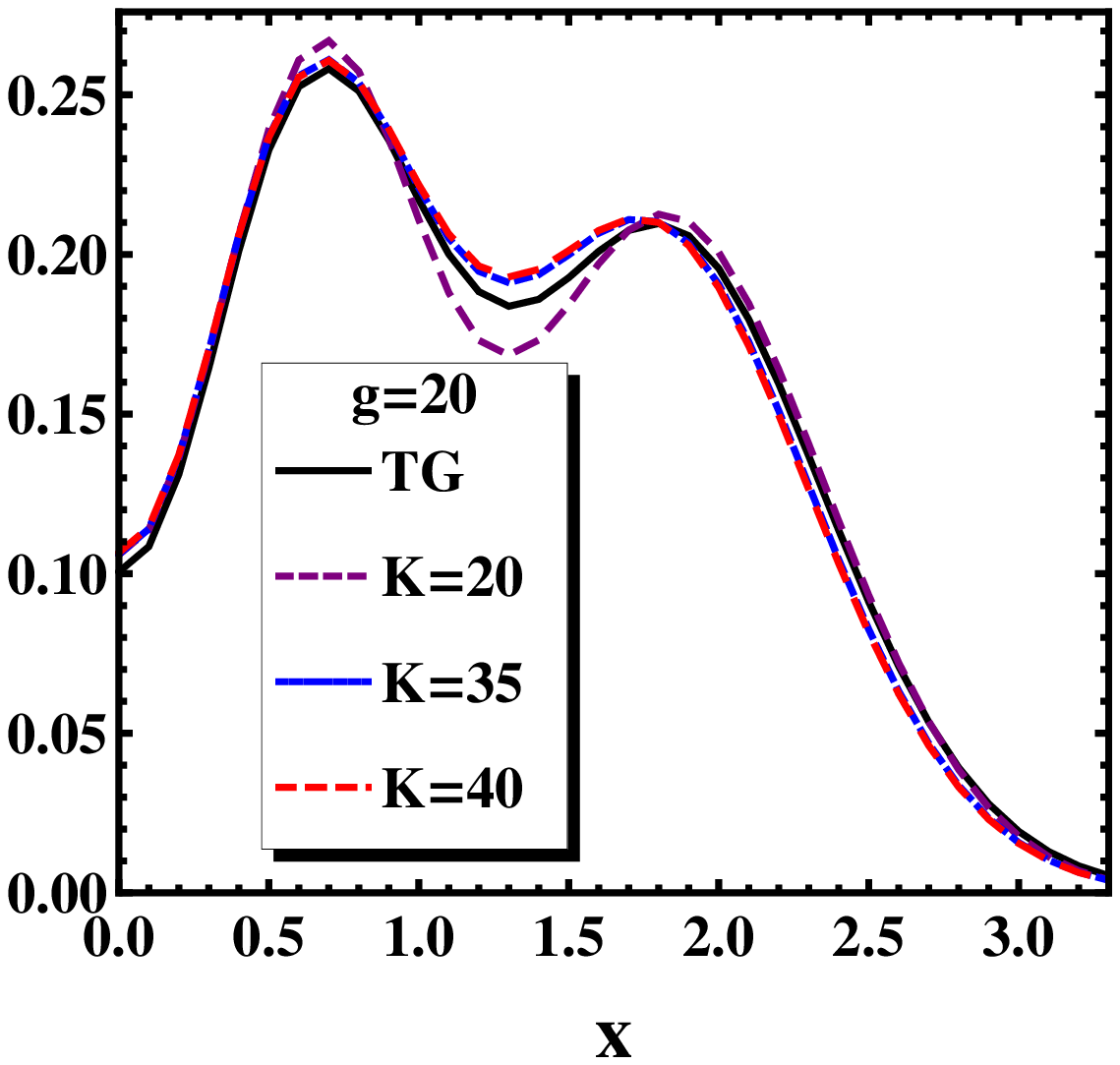}
\end{center}
\caption{\label{Fig3}
Profiles of the one-body densities for the four-particle system  in the presence of the trapping potential (\ref{exp2}), with $h=4$ and $\delta=0.2$, obtained for different values of the interaction strength $g$ and the   cut-off $K$.}
\end{figure}

Finally, we shed some light on the applicability of the OED approach with HO eigenfunctions to systems with trapping potential that is different from the harmonic trap. Here, we consider a model of a double-well potential of the form   \begin{equation}\label{exp2}
{ V}(x)={1\over 2}x^2+h \mathrm{e}^{-{x^2\over 2\delta^2}},
\end{equation}
 which is often used in literature to simulate different physical situations.
  To illustrate this, we demonstrate the convergence of the OED method for the example of a four-particle system with  control parameter
 values: $h=4$ and $\delta=0.2$. Figure \ref{Fig3} shows the ground-state one-body densities for $g=4$ and  $g=20$
that are obtained with the OED method   for  different  cut-off values of $K$.
   In the  case of $g=20$, the system  enters the TG regime,
    and in order to confirm the validity of our calculations the results  are shown
    along with the \textit{exact} solution in the TG limit as $g\rightarrow \infty$, $\rho_{TG}(x)=1/4\sum_{i=0}^{3} \phi_{i}^2(x)$ \cite{Girardeau}, where $\phi_{i}$ are the solutions of the one-particle Hamiltonian with (\ref{exp2}). Indeed, as can be seen, the limit of infinite interaction is almost reached at  $g=20$.  Clearly, in both cases that are considered,  well-converged density profiles are obtained with moderate numbers of many-body basis functions.  These results are very promising and allow us to expect that   the ED method   with  HO eigenfunctions optimized by $\Omega$ may also be an effective tool  for handling systems    with   more complex forms of trapping potentials, such as  multi-well  potentials.

 \section{Conclusions}\label{con}In conclusion,   the ED  method with an optimized  basis of  harmonic oscillator eigenfunctions has been tested on  systems with delta interactions.
Our results shows that minimization of the eigenenergy of the desired level with respect to the frequency parameter greatly improves the accuracy of the resulting approximate energy and also that of its wave function. Careful testing of the optimized ED method    for harmonic and double-well   potentials has  proved its efficiency even in the regime of  strong correlations. The use of the optimization scheme for other systems with delta interactions that are currently under intensive study,  such as quantum mixtures, is a straightforward task which could also lead to many benefits.


\bibliographystyle{abbrv}


\end{document}